\documentclass[aps,twocolumn]{article}
\usepackage{amsbsy,amssymb}
\usepackage{epsfig}
\usepackage{graphicx}
\newcommand{\be}{\begin{equation}}
\newcommand{\ee}{\end{equation}}
\newcommand{\bea}{\begin{eqnarray}}
\newcommand{\eea}{\end{eqnarray}}
\newcommand{\ba}{\begin{array}}
\newcommand{\ea}{\end{array}}
\newcommand{\E}{{\sf E}}

\title{Bias Analysis in Entropy Estimation}

\author{Thomas Sch\"urmann\\
Research, Westdeutsche Genossenschafts-Zentralbank eG,\\ 
Ludwig-Erhard-Allee 20, 40227 D\"usseldorf, \\Germany \\
\\Email: thomas.schuermann@vr.wgz-bank.de
}

\date{\today }

\begin{document}
\maketitle

{\bf Abstract:} We consider the problem of finite sample corrections for entropy estimation. New estimates of the Shannon entropy are proposed and their systematic error (the bias) is computed analytically. We find that our results cover correction formulas of current entropy estimates recently discussed in literature. The trade-off between bias reduction and the increase of the corresponding statistical error is analyzed.
\\

Statistical fluctuations of small samples induce both statistical and 
systematic deviations of entropy estimates. In the naive ("likelihood") estimator one replaces the discrete probabilities $p_i$, for $i=1,...,M$, in the Shannon entropy \cite{shann}
\be\label{Entropie}
H=-\sum\limits_{i=1}^{M}\,p_i\ln p_i,			\label{shannon}
\ee
by maximum likelihood estimates $\hat{p}_i$. More precisely, we 
consider samples of $N$ observations, and let $n_i$ be the frequency of 
realization $i$ in the ensemble. Then, with the choice $\hat{p}_i=\frac{\,n_i}{N}$, 
the naive estimate
\be
\hat{H}=-\sum\limits_{i=1}^{M}\,\hat{p}_i \ln \hat{p}_i,\label{naiv}
\ee
leads to a systematic underestimation of the entropy $H$. \\
There is a series of publications trying to improve the estimation error successively with suitable terms of corrections. One approach is to apply a 
Taylor expansion around the probability $p_i$ to the $\ln$-function in 
(\ref{naiv}) \cite{miller,harris,herzel}. A detailed computation of the expectation value of $\hat{H}$ with respect to the multinomial distribution
\be
\rho(n_1,...,n_M;p_1,...,p_M,N)= N!\prod_{i=1}^M\,\frac{p_i^{n_i}}{n_i}, 
	\label{multi}
\ee
up to the second order in $N$ was given by Harris \cite{harris} and gives
\be
\E[\hat{H}]=H-\frac{M-1}{2N}+\frac{1}{12N^2}\left(1-
\sum_{i=1}^M\frac{1}{p_i}\right)+\mathcal{O}(N^{-3}).		\label{Taylor}
\ee
The $\mathcal{O}(1/N)$ correction term was first obtained by Miller \cite{miller}. The term of order $1/N^2$ involves the unknown probabilities $p_i$, and can not be generally estimated reliably. In particular, it would not be sufficient to replace them by $\hat p_i$ in this term. \\

In order to extend the estimation beyond corrections of order $1/N$, Paninski \cite{paninski} applies {\it Bernstein approximating polynomials}, which are defined as a linear combination of binomial polynomials. It can be shown, using results from approximation theory, that there exist expansion coefficients such that the maximum (over all $p_i$) systematic deviations are of the order $1/N^2$. This is better than the order $1/N$ rate offered by the correction terms mentioned above. Unfortunately, the good approximation properties of this estimator are a result of a delicate balancing of large, oscillating coefficients, and the variance of the corresponding estimator turns out to be very large \cite{paninski}. Thus, to find a good estimator, one has to minimize bounds on bias and variance simultaneously. The result is a regularized least-squares problem, whose closed-form solution is well known. However, one can only hope that the solution of the regularized problem implies a good polynomial approximation of the entropy function. The latter also depends on whether the experimenter is more interested in reducing bias than variance, or vice versa. \\  

An alternative approach, where only observables appear in the correction term, was proposed by Grassberger \cite{grass88}. There it was assumed that all $p_i\ll 1$, so that each $n_i$ is a random variable which should follow a Poisson distribution. To start with, we consider Renyi entropies of order $q\ge 0$
\be
H(q)=\frac{1}{1-q}\, \ln\,\sum\limits_{i=1}^{M}\,p_i^q.	\label{Renyi}
\ee
The Shannon case results from taking the limit $q\to 1$, 
i.e. $H=\lim_{q\to 1} H(q)$. For the estimation of $H(q)$ it seems obvious first to ask for an unbiased estimator of any term $p_i^q$ of the sum in (\ref{Renyi}). In the case of integer values of $q\ne 1$ the situation is trivial because the unique unbiased estimator $\widehat{p^q}$ is\footnote{For simplification the index $i$ will be omitted.} 
\be
\widehat{p^q} = \frac{1}{N^q}\,\frac{n!}{(n-q)!}\qquad n\ge q,  
\label{poiss}
\ee
with $\widehat{p^q}:=0$ for $n<q$. However, to achieve $q\to 1$, it is necessary to look first for a generalization for arbitrary $q$. As shown in \cite{grass88}, the analytical continuation of the estimator is non-trivial since a naive replacement of the factorials in (\ref{poiss}) by $\Gamma$-functions is biased. Indeed, unbiased estimators of $p^q$ do not exist for non-integer values of $q$. Nevertheless, in \cite{grass88} an interesting estimator of $p^q$ was proposed which is at least asymptotically unbiased for large $N$, and is also a "good" approximation in the case of small samples. The corresponding estimator of the Shannon entropy is\footnote{The summation is defined for all $n_i>0$. The digamma function $\psi(n)$ is the logarithmic derivative of the $\Gamma$-function, see e.g. \cite{abramow}} \cite{grass88}
\be
\hat{H}_\psi = \sum_{i=1}^M\,\frac{n_i}{N}\,\left(\ln N-\psi(n_i)-
\frac{\,(-1)^{n_i}}{n_i(n_i+1)}\right).				\label{psi}
\ee
For the interesting case of small probabilities $p_i\ll 1$ the estimate (\ref{psi}) is less biased than the estimator obtained by the Miller correction. \\
A further improvement, related to the latter approach, which is also based on the assumption of Poisson distributed frequencies, was recently proposed by Grassberger \cite{grass03}. The corresponding entropy estimator of the Shannon entropy is 
\be
\hat{H}_G=\sum_{i=1}^M\, \frac{\,n_i}{N}\left(\psi(N)-\psi(n_i)-(-
1)^{n_i}\int_0^1\frac{\;t^{n_i-1}}{1+t}dt\right).		\label{Gn}
\ee
The correction term of the earlier estimator $\hat{H}_\psi$, is recovered by a series expansion of the integrand in (\ref{Gn}) up to the second order. The higher order terms of the integrand lead to successive bias reductions compared to (\ref{psi}). 

At this point, one might ask whether further improvements in bias reduction are possible. Moreover, it is of special interest to consider the trade-off between bias reduction and the increase of the corresponding statistical error. In the following theorem, we propose a family of new entropy estimators and determine their systematical error analytically. We will present a detailed analysis of the bias and show that the entropy estimators above are specific examples of our general results. 

In view of the following computations we note that the Shannon entropy is a sum of terms, $h(p_i)=-p_i\ln p_i$, which exclusively depend on the class $i$, for $i=1,...,M$. Therefore, when we consider expectation values with respect to $n_i$, the computations can be carried out by replacing the joint distribution (\ref{multi}) by the binomial distribution
\be
   P(n_i;p_i,N) = {N\choose n_i} p_i^{n_i} (1-p_i)^{N-n_i},	    
\label{binomi}
\ee
for $0\le n_i\le N$ and $\E[n_i] = p_iN$. Now let us consider the following\\
\\
{\bf Theorem:} Let $\xi>0$ be a real number and 
\be
\hat{h}(\xi,n) = \frac{\,n}{N}\,\bigg(\psi(N)-\psi(n)
- (-1)^n\int\limits_0^{1/\xi-1}\frac{\;t^{n-1}}{1+t}\,dt\, \bigg), 
\label{esti}
\ee
be a parametric family of estimators of the function $h(p)=-p\,\ln p$. For the particular case $n=0$, let $\hat{h}(\xi,0)=0$. Then, we have the identity 
\be
\E[\hat{h}(\xi,n)] = -p\ln p + b(\xi,p),	\label{id}
\ee
and 
\be
b(\xi,p)= -p\,\int_0^{1-p/\xi}\frac{\;\;t^{N-1}}{1-t}\;dt	\label{b}
\ee
is the bias of the estimator $\hat{h}(\xi,n)$.\\
\\ 
From the theorem we directly obtain an estimator of the Shannon entropy by summation of (\ref{esti}), i.e. $\hat{H}_S(\xi)=\sum_{i=1}^M\hat{h}(\xi,n_i)$. Using a similar notation as in \cite{grass03} we receive the following expression
\be
\hat{H}_S(\xi)=\psi(N)-\frac{1}{N}\sum_{i=1}^M\, n_i\,S_{n_i}(\xi) 
\label{Hs}
\ee
with
\be
S_n(\xi)=\psi(n)+ (-1)^{n}\int_0^{1/\xi-1}\frac{\;t^{n-1}}{1+t}\,dt.
\label{Sn}
\ee
{\bf Proof:} For real $q\ge 0$ we consider the finite Taylor series approximation of $p^q$ around $\xi>0$, i.e.
\be
T_N(p) = \sum_{\nu=0}^N {q\choose \nu}\xi^{q-\nu}(p-\xi)^\nu,	\label{reihe}
\ee
with
\be\label{koeff}
{q\choose\nu}=
\prod\limits^{\nu}_{i=1}\,\frac{q-i+1}{i}.
\ee
We expand the brackets on the right hand side of (\ref{reihe}) and rearrange the terms in order to obtain the following double summation
\be
T_N(p) = \sum_{n=0}^N (-1)^n p^n  \xi^{q-n} \sum_{\nu=n}^N {q\choose 
\nu}{\nu\choose n} (-1)^\nu.
\ee
For simplification we introduce the substitution
\be
F(\xi,q,k,N) = (-1)^k{\xi}^{q-k} \sum_{\nu=k}^N {q\choose \nu}{\nu\choose 
k} (-1)^{\nu}.
\ee
Then, by further algebraic manipulations we obtain the identity
\be
\frac{T_N(p)}{(1-p)^N} = \sum_{n=0}^N \Theta^n \sum_{k=0}^n {N-k\choose n-k} 
F(\xi,q,k,N) \label{SN}
\ee
with $\Theta=p/(1-p)$. The rhs of the latter expression is a polynomial in $\Theta$ whose $(N+1)$ coefficients are all independent of the probability $p$. On the other hand, there is an unbiased estimator, say $\hat\delta_q(\xi,n)$, of the expansion $T_N(p)$, since the expectation value of $\hat\delta_q(\xi,n)$ can also be expressed by a polynomial of finite order in $\Theta$, i.e.
\be
\frac{\E[\hat\delta_q(\xi,n)]}{(1-p)^N} = \sum_{n=0}^N {N\choose 
n}\;\hat\delta_q(\xi,n)\; \Theta^n. \label{delta}
\ee
To obtain the explicit expression of $\hat\delta_q(\xi,n)$, we consider the necessary/sufficient condition for unbiasedness 
\be
\E[\hat\delta_q(\xi,n)]= T_N(p).				\label{estgl}
\ee
After inserting (\ref{SN}) and (\ref{delta}) into to the condition 
(\ref{estgl}) and then comparing coefficients, it follows
\be
\hat\delta_q(\xi,n)= {N\choose n}^{-1}\sum_{k=0}^n F(\xi,q,k,N) {N-k\choose 
n-k}.  \label{delta1}
\ee
This unbiased estimator of $T_N(p)$ is unique because the identity (\ref{estgl}) is satisfied for arbitrary $\Theta$. Next we carry out the derivation of $\hat\delta_q(\xi,n)$ with respect to $q$, and consider the limes $q\to 1$. For this purpose we note that the derivative of the binomial coefficient (\ref{koeff}) is 
\be
\frac{d}{dq}{q\choose \nu}\bigg|_{\,q=1}=
\left\{ \begin{array}{r@{\quad\quad}l}
(-1)^\nu \frac{1}{ \nu(\nu-1) } & \nu\ge 2\\
\nu \qquad & \nu=0,1. \label{abl}
\end{array} \right.
\ee
By direct computation it follows, that the negative derivative of the estimator $\hat\delta_q$, for $q\to 1$, is given by the expression
\bea
-\lim\limits_{q\to1}\frac{\,d\hat\delta_q}{dq} &=& \frac{n}{N}\left(\psi(N)-\psi(n)\right)+\frac{\xi}{N}\left(1-\frac{1}{\xi}\right)^n \nonumber\\
&-&\frac{n}{N} (-1)^n\int\limits_0^{1/\xi-1}\frac{\;t^{n-1}}{1+t}\,dt. 
\eea
On the other hand, when applying the same procedure to the Taylor series expansion $T_N(p)$, we find 
\bea
-\lim\limits_{q\to1}\frac{\,d T_N}{dq} &=& -p\ln p + \frac{\xi}{N}\left(1-\frac{p}{\xi}\right)^N \nonumber\\
&-& p \int\limits_0^{1-p/\xi}\frac{\;t^{N-1}}{1-t}\,dt.
\eea
Equating both by using (\ref{estgl}) and applying the trivial identity $\E[(1-1/\xi)^n]\equiv(1-p/\xi)^N$, by using the notation of the theorem, we obtain the result 
\be
\E[\hat{h}(\xi,n)] = -p\ln p + b(\xi,p).	\label{result}
\ee
Thus, the claim (\ref{id}) has been proven. Finally, we consider the residual term, $R_{N+1}$, of the Taylor series expansion $T_N(p)$. By definition, the identity $p^q=T_N(p)+R_{N+1}(p)$ is valid. Using the latter and applying the ordinary integral representation of $R_{N+1}$, then we find the following relation between the bias and the first derivative of the residual term 
\bea
\lim\limits_{q\to 1} \frac{\,dR_{N+1}}{dq} = b(\xi,p) + \frac{\xi}{N}\left(1-\frac{p}{\xi}\right)^N. 
\eea
$\hfill\Box$\\
\\
\begin{figure}
\begin{center}
\psfig{file=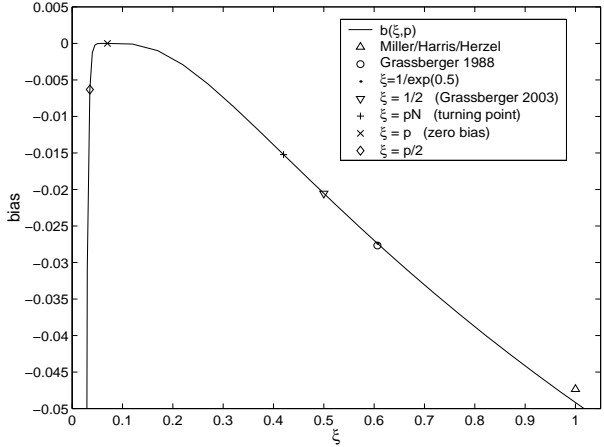,width=8.0cm, height=6.0cm }
\caption{
Systematic error of $\hat{h}(\xi,n)$ for samples of $N=6$ observations and $p=0.07$. Several special cases of $\xi$ are shown. The case $\xi=e^{-\frac{1}{2}}$ slightly improves the estimator $\hat{H}_\psi$ (see the dot within the circle).  
\label{fig1}} \end{center}
\end{figure}
Every point on the continuous line in Fig.\ref{fig1} is the bias of the corresponding estimator $\hat{h}(\xi,n)$. It is unbiased for $\xi=p$, and there is a turning point for $\xi=pN$. The estimator is {\it asymptotically unbiased}, i.e. $b(\xi,p)\to 0$ for $N\to\infty$, if $\xi\ge p/2$. On the other hand, in Fig.\ref{Variance} we see the mean square error (statistical error) $\sigma^2(\xi,p)=\E[(\hat{h}(\xi,n)-h(p))^2]$. The trade-off between bias and the statistical error of the estimator is shown in Fig.\ref{bV}. Typically, one is more interested in the error of the entire sum over the states $i$ in Eq.(\ref{Entropie}). If there are $M$ terms, and if each is roughly of the same order of magnitude, then the total bias and the total variance are both ${\sim M}$, thus the statistical deviation increases only as $M^{1/2}$. Therefore, the more terms one has (the larger $M$), the more one is interested in using small values of $\xi\geq p/2$, if one wants the total statistical and the total systematic deviations to have the same size. Thus, the interesting estimators lie between both extremes, i.e. the minimum statistical error, and in case $\xi=p$ with vanishing bias. The following particular cases are especially interesting to focus on:

$\underline{\xi=1}$: In this case we obtain the trivial estimator for $h(p)$
\begin{figure}
\begin{center}
\psfig{file=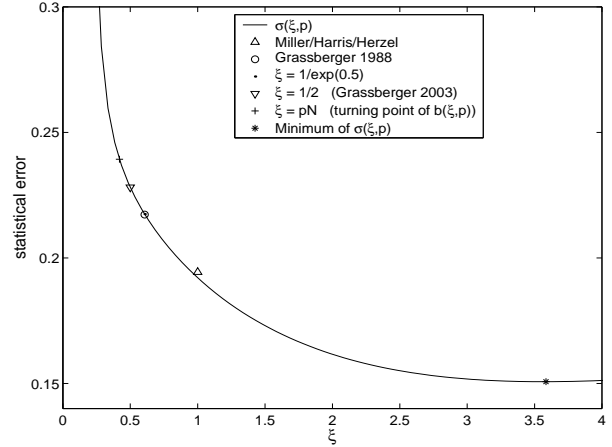,width=8.0cm, height=6.0cm }
\caption{
	Numerical computation of the statistical error 	$\sigma(\xi,p)$ for 	samples of $N=6$ 	observations and $p=0.07$. The minimum of 	$\sigma(\xi,p)$ is obtained by the solution of $\E[\hat h\frac{\partial 	\hat h}{\partial \xi}] = h\frac{\partial b}{\partial\xi}$. 
\label{Variance}} \end{center}
\end{figure}
\begin{figure}
\begin{center}
\psfig{file=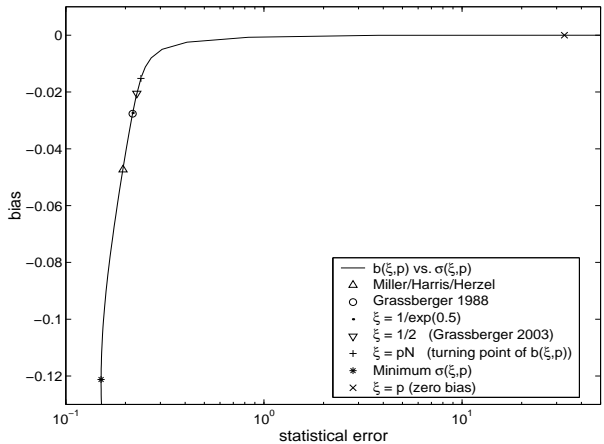,width=8.0cm, height=6.0cm }
\caption{
	Trade-off between bias and statistical error for samples of $N=6$ and 	$p=0.07$. The reduction of the bias for $\xi<pN$ is related to a 	strongly increasing statistical error. On the 	other hand, the bias 	corresponding to the minimum statistical error is larger than all the 	above mentioned estimators.
\label{bV}} \end{center}
\end{figure}
\be
\hat{h}(1,n) = \frac{\,n}{N}\,(\psi(N)-\psi(n)),
\ee
and $\hat{h}(1,0)=0$ for $n=0$. By the identity (\ref{id}) we receive the following expectation value 
\be 
\E[\hat{h}(1,n)]= -p\ln p -p\,\int_0^{1-p}\frac{\;\;t^{N-1}}{1-t}\;dt. 
\label{b1}
\ee
The latter expression has been recently mentioned in \cite{grass03} 
(citation [14] in it). In the asymptotic regime $n\gg 1$ it leads to the Miller correction\footnote{This is because in the asymptotic regime we have the relation $\psi(x)\sim\ln(x)-1/2x$.}, i.e. $\hat{h}(1,n)= -\frac{n}{N}\ln \frac{n}{N}+\frac{1}{2N}+{\cal O}(1/N^2)$. 

$\underline{\xi= e^{-\frac{1}{2}}}$: The Grassberger estimator $\hat{H}_\psi$ is a special case, since it is not exactly covered by our theorem. However, it can be very well approximated, if the Taylor expansion is chosen around the particular value of $\xi=e^{-\frac{1}{2}}$. By numerical analysis we verified that the corresponding estimator $\hat{h}(e^{-\frac{1}{2}},n)$ is less biased than the estimator (\ref{psi}), for any $N>1$ and arbitrary $p$. In Fig.\ref{fig1}, we see that there is almost no difference between both estimators. However, by numerical verification, slight improvements become visible for larger probabilities (e.g. $p>0.8$). In the case of a single observation, i.e. $N=1$, there is no difference between the two, for any $p$.

$\underline{\xi=\frac{1}{2}}$: This case is identical to the Grassberger estimator (\ref{Gn}), see \cite{ grass03}. As shown in Fig.\ref{fig1}, it is less biased than the Miller estimator and the estimator $\hat{h}(e^{-\frac{1}{2}},n)$. But the statistical error of $\hat{h}(\frac{1}{2},n)$ is slightly bigger as we can see in Fig.\ref{Variance}. In the left half of the unit interval, i.e. $\xi<\frac{1}{2}$, we obtain further reduction of the bias. But now one has to be attentive since we have the limes $|\hat{h}(\xi,n)|\to\infty$ for $n\to\infty$. Although $n$ is always finite in practice, this behavior is an indication that the statistical error of all estimators with $\xi\ll 1$ could increase very fast. The dramatic increase of $\sigma(\xi,p)$ for $\xi\to 0$ is shown in Fig.\ref{bV}. Therefore, the particular choice $\xi=\frac{1}{2}$ seems to be very suitable for estimation because it has the smallest $|b(\xi,p)|$ with $\hat h\to h(p)$ for $n\to\infty$ and any $p\in (0,1]$.
On the other hand, the most conservative case is given by the minimum variance estimator (see Fig.\ref{bV}). In this case the value of the statistical error and the absolute value of the bias are comparable. A compromise between both extremes might be the estimator for $\xi= e^{-\frac{1}{2}}\approx 0.6$. This case is less biased than the minimum variance estimator, and less risky than the Grassberger estimator $\hat H_G$.\\
\\
To sum up, in the above analysis, we see that it is not possible to decide which of the many estimators $\hat{h}(\xi,n)$ should be generally preferred. A good choice of the parameter $\xi$ always depends on the special application under consideration and the individual preference of the scientist.  



\begin{thebibliography}{99}
\bibitem{shann} C. E. Shannon and W. Weaver, {\it The Mathematical Theory of Communication}, (University of Illinois Press, Urbana, IL 1949).
\bibitem{miller} G. Miller, Note on the bias of information estimates.
   In H. Quastler, ed., {\it Information theory in psychology II-B}, 
   pp 95-100 (Free Press, Glencoe, IL 1955).
\bibitem{harris} B. Harris, Colloquia Math. Soc. Janos Bolyai, p. 323 
(1975).
\bibitem{herzel} H. Herzel, Sys. Anal. Mod. Sim. {\bf 5}, 435 (1988).
\bibitem{paninski} L. Paninski, Neural Computation {\bf 15}, 1191 (2003).
\bibitem{grass88} P. Grassberger, Phys. Lett. {\bf A 128}, 369 (1988).
\bibitem{grass03} P. Grassberger, www.arxiv.org, physics/0307138 (2003).
\bibitem{abramow} M. Abramowitz and I. Stegun, eds., {\it 
   Handbook of Mathematical Functions} (Dover, New York 1965).
\end{thebibliography}
\end{document}